\documentclass[sigconf]{acmart}

\usepackage[inline]{enumitem}

\usepackage{amsmath}
\DeclareMathAlphabet{\mathcal}{OMS}{cmsy}{m}{n}
\usepackage{csquotes}
\usepackage{balance}

\AtBeginDocument{%
  \providecommand\BibTeX{{%
    \normalfont B\kern-0.5em{\scshape i\kern-0.25em b}\kern-0.8em\TeX}}}

\copyrightyear{2022}
\acmYear{2022}
\setcopyright{acmcopyright}\acmConference[UMAP '22]{Proceedings of the 30th ACM Conference on User Modeling, Adaptation and Personalization}{July 4--7, 2022}{Barcelona, Spain}
\acmBooktitle{Proceedings of the 30th ACM Conference on User Modeling, Adaptation and Personalization (UMAP '22), July 4--7, 2022, Barcelona, Spain}
\acmPrice{15.00}
\acmDOI{10.1145/3503252.3531314}
\acmISBN{978-1-4503-9207-5/22/07}

\begin{document}

\title{Critiquing-based Modeling of Subjective Preferences}

\author{Alan Medlar}
\affiliation{%
  \institution{University of Helsinki}
  \streetaddress{Helsinki}
  \country{Finland}
}
\email{alan.j.medlar@helsinki.fi}

\author{Jing Li}
\affiliation{%
  \institution{University of Helsinki}
  \streetaddress{Helsinki}
  \country{Finland}
}
\email{jing.li@helsinki.fi}

\author{Yang Liu}
\affiliation{%
  \institution{University of Helsinki}
  \streetaddress{Helsinki}
  \country{Finland}
}
\email{yang.liu@helsinki.fi}

\author{Dorota G{\l}owacka}
\affiliation{%
  \institution{University of Helsinki}
  \streetaddress{Helsinki}
  \country{Finland}
}
\email{dorota.glowacka@helsinki.fi}

\begin{abstract}
Applications designed for entertainment and other non-instrumental purposes are challenging to optimize because the relationships between system parameters and user experience can be unclear. Ideally, we would crowdsource these design questions, but existing approaches are geared towards evaluation or ranking discrete choices and not for optimizing over continuous parameter spaces. 
In addition, users are accustomed to informally expressing opinions about experiences as critiques (e.g.~it's too cold, too spicy, too big), rather than giving precise feedback as an optimization algorithm would require.
Unfortunately, it can be difficult to analyze qualitative feedback, especially in the context of quantitative modeling.

In this article, we present \textit{collective criticism}, a critiquing-based approach for modeling relationships between system parameters and subjective preferences. We transform critiques, such as ``it was too easy/too challenging'', into censored intervals and analyze them using interval regression. Collective criticism has several advantages over other approaches: ``too much/too little''-style feedback is intuitive for users and allows us to build predictive models for the optimal parameterization of the variables being critiqued. We present two studies where we model: 
\begin{enumerate*}[label=(\roman*)]
\item aesthetic preferences for images generated with neural style transfer, and 
\item users' experiences of challenge in the video game Tetris. 
\end{enumerate*}
These studies demonstrate the flexibility of our approach, and show that it produces robust results that are straightforward to interpret and inline with users' stated preferences. 

\end{abstract}

\ccsdesc[500]{Human-centered computing~User models}
\ccsdesc[500]{Human-centered computing~User studies}

\keywords{critiquing, subjective preferences, user experience, randomized experiments}

\maketitle

\section{Introduction}

User experience is an increasingly important consideration in systems, services and products, with many applications emphasizing various aesthetic, affective and hedonic properties \cite{hassenzahl2006user,hassenzahl2008user}. Such qualities are particularly important in applications with no instrumental purpose, such as video games, digital toys and interactive artwork \cite{swink2008game, bernhaupt2010user, sanchez2012playability}. 
Unfortunately, these applications can be challenging to optimize as they can contain numerous parameters that impact user experience, but for which there are no objective quantities to be maximized and, therefore, need to be set on the basis of subjective preferences. 
%
In video games, for example, character movement speed has a significant impact on user experience: too slow and navigating the game world can be tedious, but too fast and it can become difficult to control \cite{swink2008game}. The sweet spot that balances these two extremes will be determined by subjective preferences that are influenced by contextual factors (e.g.~larger characters tend to move slower) and other human factors (e.g.~expert players might prefer a faster, more challenging game) \cite{swink2008game}. 

How should we go about setting such parameters? 
A single user (e.g.~the developer) could simply set a given parameter to whatever feels intuitively correct. 
Alternatively, multiple users could be recruited to provide 
feedback via questionnaires or to state their preference for one parameterization over another (i.e.~pairwise comparisons).
%
Lastly, we could make assumptions about how user behavior relates to experience, e.g.~we could assume that engagement (time spent) correlates with positive user experiences. 
Unfortunately, all these approaches have downsides to inferring subjective preferences: 
a single user is unlikely to be representative of the group, 
questionnaire response scales need to be validated \cite{lindgaard2013introduction} and require large sample sizes per parameterization. 
Pairwise comparisons do not scale, with the number of comparisons increasing exponentially with the input size \cite{perez2017practical}.
Finally, behavioral data can be misinterpreted; correlating with negative experiences as well as positive ones \cite{tufekci2018youtube}. 

In this article, we present a critiquing-based approach for modeling relationships between system parameters and subjective preferences called {\em collective criticism}. We consider a crowdsourcing scenario where a group of individuals are presented with a system or product and asked to critique their experiences in terms of a given attribute. 
These critiques take the form of retrospective summaries, such as ``the food was too spicy/bland'' after eating a meal, or ``the weather was too hot/cold'' when describing a vacation \cite{chen2012critiquing, medlar2017towards}. 
Such feedback is qualitative, but we can extract quantitative information by considering the context within which the feedback was given. 
Using the video game example from before, if character movement is ``too slow'' and the speed was set to $x$, then this implies that the optimal speed lies in the interval from $x$ to $+\infty$ (strictly speaking, the right-censored interval $(x,+\infty]$). Importantly, users do not need to know anything about how the game was implemented to give their critiques: they are based solely on their experiences. 
This method of transforming critiques into intervals allows us to model relationships between user experience and system parameters using interval regression. 
We demonstrate the generality of this approach using two studies related to images generated using neural style transfer and users' experience of challenge after playing the video game Tetris.

At present, our approach cannot guard against abuse from bots and disinterested participants on crowdsourcing platforms like Amazon Mechanical Turk. Instead, we focused on groups of users that may not have any particular expertise, but are assumed to provide honest feedback. Examples of such groups include beta testers for a video game that is currently under development or a design community that meets to critique one another's work. To replicate these scenarios, we present studies performed in uncontrolled environments where participants were informally recruited by interrupting their daily lives, e.g. during a conference coffee break or while studying in the university library. 
%
%
The main contributions of this paper are as follows:
\begin{itemize}[leftmargin=0.2in,itemsep=4pt,topsep=6pt,parsep=0pt,partopsep=0pt]

    \item A novel critiquing-based approach for modeling relationships between system parameters and subject preferences called \emph{collective criticism} that combines randomized tasks with summary retrospective feedback.
    
    \item A modeling approach that transforms critiques into censored intervals to be 
    analyzed using statistical packages for interval regression.
    
    \item We present 
    two case studies using collective criticism to model 
    (i) aesthetic preferences for images generated using neural style transfer, and 
    (ii) user experiences of challenge after playing the video game Tetris. 
    
\end{itemize}

\section{Related Work}

In this section, we review related work on crowdsourcing preferences and 
how critiquing has been used as an interaction mechanism in different types of information systems.

\subsection{Crowdsourced Preferences}

Crowdsourcing has eased the collection of subjective preferences by reducing cost and turnaround time,
while showing a high degree of agreement with data collected in controlled laboratory environments \cite{behrend2011viability,crump2013evaluating,tse2016crowdsourcing}.
In computing, crowdsourcing has been used to collect subjective preferences for algorithm development and evaluation.
In particular, it has been used to collect 
relevance judgments \cite{alonso2008crowdsourcing},
data for sentiment analysis \cite{bakshi2016opinion},
assessments of toxicity in online discussions \cite{aroyo2019crowdsourcing} and even 
examples of irony and sarcasm \cite{filatova2012irony}.
In these areas, items are scored independently of one another 
using either binary (e.g.~relevant/not relevant) or ordinal (e.g.~negative to positive) labels.
While these labels are considered subjective, there is assumed to be a consensus within a given culture or community. 
As this data tends not to be aggregated, but used as training data, each label needs to be correct 
and, therefore, there is an extensive literature on study design and statistical methods for quality control 
(for a recent survey, see \cite{jin2020technical}).
%

In other domains, crowdsourcing is used to understand the opinions and aesthetic preferences of the general public.
It has been used, for example, to study the aesthetics of 
platforming games 
\cite{shaker2012crowdsourcing}, 
3D models \cite{dev2017polygons} and 
portrait photography \cite{expressions33mirror}.
Crowdsourcing aesthetic preferences has seen extensive use in reconstructive and cosmetic surgery.
In reconstructive surgery, it has been used to compare the aesthetics of cleft lip outcomes \cite{tse2016crowdsourcing} 
and the results of different surgical techniques \cite{suchyta2020applied}. 
Whereas, in cosmetic surgery, it has been used 
to assess buttock augmentation outcomes 
\cite{vartanian2018ideal} and
to characterize anatomical aesthetic preferences for male \cite{massie2021defining} and female \cite{frojo2021defining} genitalia.
The use of crowdsourcing is considered important as 
surgical aesthetic outcomes are usually only assessed by an individual, either the patient or surgeon, 
which can lead to biased assessments \cite{azadgoli2019public}. 
In this article, we include a study comparing the aesthetic preferences of an individual (in our case, a developer) with our  approach and find similar disparities.
Beyond aesthetics, crowdsourcing has recently been used to understand the public perception of topical subjects, 
such as AI fairness \cite{van2019crowdsourcing} and moral decision-making in the context of autonomous driving \cite{awad2018moral}. 
All of the above examples used either questionnaires or pairwise comparisons to infer subjective preferences, both of which have weaknesses. Questionnaire response scales need to be validated to ensure they are measuring what they purport to measure \cite{lindgaard2013introduction}. Pairwise comparisons do not suffer from this issue, but can require exceptionally large sample sizes (i.e.~multiples of ${n \choose 2}$ pairwise comparisons), practically limiting assessment to a relatively small number of items \cite{perez2017practical}. 
Our approach does not require validation like a questionnaire because we only ask a single question that is often directly referencing a given parameter (i.e.~we do not use multi-item scales, which would require us to assess construct validity).  
Furthermore, as users are critiquing individual parameters and not making pairwise comparisons, sample sizes can be much lower.


While not usually considered crowdsourcing, A/B testing uses randomized experiments to compare the effectiveness of two (or more) versions of the same system on the basis of conversion rates, e.g.~click-throughs or purchases \cite{kohavi2017online}. This makes the assumption that a given behavior is correlated with positive user experience \cite{kohavi2017surprising}. However, it can also result in unintended consequences if the measured response corresponds to multiple outcomes, e.g.~video watch time is correlated with outrage as well as enjoyment \cite{tufekci2018youtube}. Our approach is based on stated preferences, avoiding the ambiguity-related issues associated with inferring revealed preferences from behavioral data.

\subsection{Critiquing}

Critiquing has previously been used as an interaction mechanism in both interactive search and conversational recommender systems \cite{chen2012critiquing}.
In critiquing-based systems, users provide critiques in relation to item features, e.g.~``too expensive'', 
to iteratively navigate a complex information space \cite{chen2012critiquing}.
The FindMe system was the first critiquing recommender, combining browsing with the critiquing of previously retrieved examples \cite{burke1997findme}.
Later systems utilized the approach to develop interactive retrieval systems for 
e-commerce \cite{burke2002interactive,faltings2004designing}, 
decision support \cite{pu2004decision} and 
preference-based search \cite{viappiani2006preference}.
However, the most common application of critiquing is in conversational recommender systems \cite{jannach2020survey},
where they have been applied to various domains, including 
movie \cite{vig2011navigating} and
music recommendation \cite{jin2019musicbot}.
Recently, critiquing-based systems have used language-based attributes, rather than fixed item features, 
to automatically identify attributes of an item that can be critiqued \cite{wu2019deep}.

In interactive systems, critiques are usually conceptualized as constraints that are applied across items \cite{burke2002interactive,faltings2004designing,pu2004decision,viappiani2006preference}. 
For example, if a plane ticket is critiqued as too expensive, it does not make sense to recommend tickets with higher prices.
%
In this work, we are concerned with modeling user preferences, so critiques are modeled probabilistically even though they may be constraints from the perspective of individuals.

\section{Collective Criticism}

Collective criticism is a critiquing-based approach for modeling the relationships between system parameters and subjective preferences. Each study based on collective criticism requires three elements:
\begin{enumerate*}[label=(\roman*)]
    \item critique elicitation,
    \item randomized tasks, and 
    \item statistical modeling
\end{enumerate*}
to analyze user preferences.

\subsection{Critique Elicitation}

During a study, each participant performs a task where they 
interact with a system or product for a short period of time. 
After completing the task, participants are asked to critique the property under study using summary retrospective feedback.
In concrete terms, summary retrospective feedback takes the form of judgments, such as ``too much'' or ``too little'' of a given property. 
For example, suppose we wanted to optimize the volume of an audible, but discreet, ringtone for an office environment. 
We would ask participants to listen to a ringtone and state whether they thought it was ``too quiet'' to be audible or ``too loud'' to be discreet.

This kind of critique is qualitative: it does not contain information related to how much something should change, merely the direction of that change. 
This allows participants to respond with their gut instinct and can be used when an appropriate scale for quantitative feedback does not exist.

\subsection{Randomized Tasks}

Here, we detail the assumptions of our approach and describe the steps necessary to design and conduct an study.

\subsubsection{Assumptions} 
We assume that investigators have a hypothesis that a given parameter, $p$, affects a specific property of the system under study. Furthermore, we assume that optimizing $p$ necessitates making a trade-off, i.e.~setting $p$ either too high or too low is detrimental to user experience, but that there is a ``sweet spot'' where the average user is maximally satisfied. 
We do not assume that there exists a single optimal parameterization for all users as this depends on the experimental design of a given study. 
Indeed, we provide a worked example in Section~\ref{sec:workedexample} that assumes a single optimal parameterization and two further case studies in Sections~\ref{sec:neuraltransfer} and \ref{sec:tetrisstudy} where $p$ is modeled in terms of other explanatory variables. 

\subsubsection{Effective Parameter Ranges}
Investigators need to determine the effective range for the parameter, $p$. This could be the entire range of the parameter, e.g.~the decision threshold in a probabilistic classifier is 0 to 1 inclusive, or be limited to a given interval. In this article, we determined the effective range of parameters by trial and error, however, it could also be limited due to physiological reasons, e.g.~human hearing is limited to 20 Hz to 20 kHz, or technological reasons, e.g.~telephony limits audio frequencies to 300 Hz to 3.4 kHz. Therefore, optimizing the frequency of a tone would have different effective ranges under different circumstances.

\subsubsection{Summary Retrospective Anchors}
Study participants give critiques using summary retrospective feedback guided by the investigator. The study design, therefore, needs to include verbal anchors to ensure that users critique the correct property. In general, verbal anchors are words used to indicate the informal meaning of response scales, such as ``strongly agree'' and ``strongly disagree''. In our case, anchors are judgments, such as ``too hot'' and ``too cold''. 

Selecting appropriate verbal anchors is important for two reasons: 
\begin{enumerate*}[label=(\roman*)]
    \item from the participants' perspective, anchors need to capture their collective understanding of the extremes of the property being assessed, and,
    \item from the investigator's perspective, anchors need to correspond to increasing and decreasing the parameter being optimized. 
\end{enumerate*}

\subsubsection{Procedure}

We extract quantitative information from summary retrospective feedback by changing the underlying conditions from which the assessment is made. We achieve this by randomizing the parameter of interest within the parameter's effective range. The procedure is as follows:
\begin{enumerate}[leftmargin=0.2in,itemsep=4pt,topsep=6pt,parsep=0pt,partopsep=0pt]

    \item We randomly set the parameter, $p$, to a value selected uniformly at random from the parameter's effective range.

    \item Participants are instructed to perform a study-specific task. Participants could be asked to use a system for a given period of time or to simply look at an image.
    
    \item After completing the task, participants are asked to critique their experience with respect to a given property using the study's summary retrospective anchors (e.g.~too high/too low).
    
    \item For each observation, we record the random value of $p$, the participant's critique and any additional metadata or user behavior data that is to be used for modeling (see Section~\ref{sec:tetrisstudy} for an example).
\end{enumerate}

\noindent If we want to understand the impact that other parameters in the system have on $p$ (i.e.~interactions), 
then these additional parameters also need to be randomized in Step 1 and recorded along with $p$ in Step 4 (see Section~\ref{sec:neuraltransfer} for an example).

\subsection{Statistical Modeling}

\begin{figure}
    \centering
    \includegraphics[width=\columnwidth]{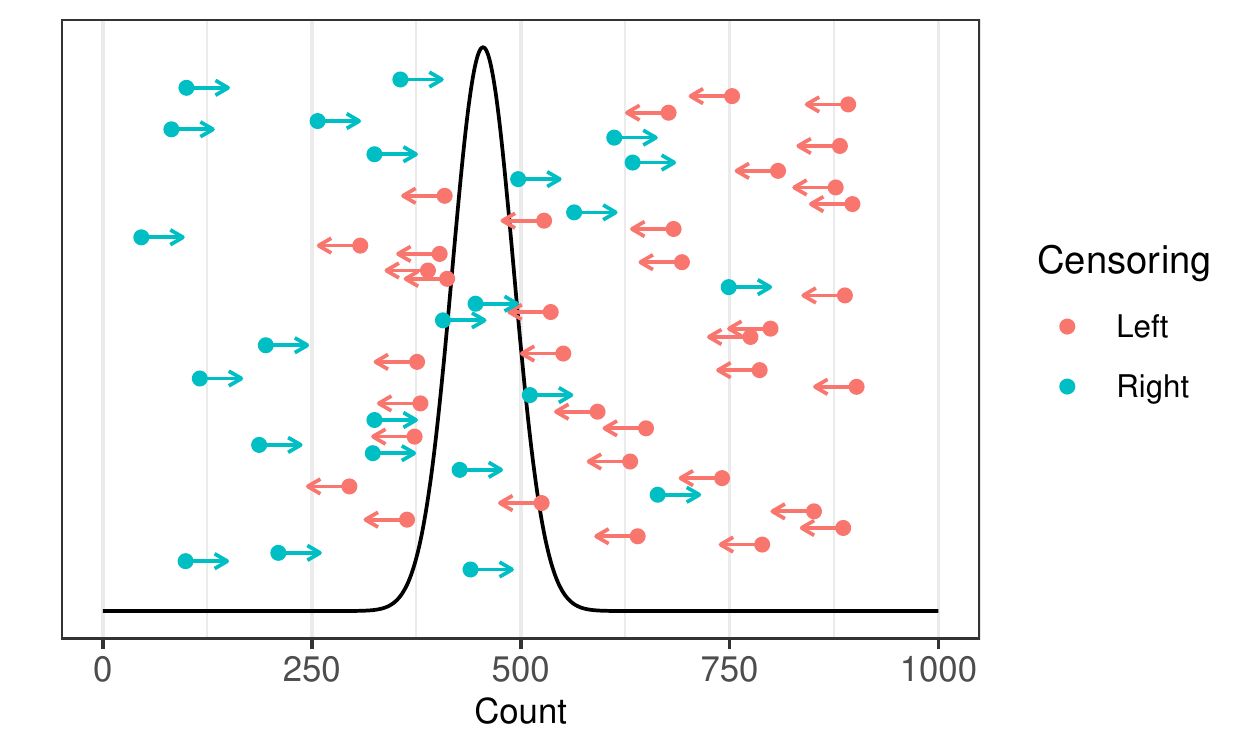}
    \caption{Interval data and final result from worked example in Section~\ref{sec:workedexample}. Left-censored intervals are colored red and right-censored intervals are blue. The $y$ coordinate of each interval is the order tasks were performed in. }
    \label{fig:intreg}
\end{figure}

After all the tasks have been performed, we model the data set using interval regression.
This requires us to transform participants' critiques into censored and/or non-censored intervals. 
We use left-censored intervals to represent when participants stated a parameter was set too high. That is, if the parameter being optimized was assigned the random value $p$, then the resulting censored interval is $(-\infty, p]$, i.e.~while we do not know the optimal value that would maximize the participant's experience, we assume that it is in the interval up to and including $p$. If the effective range of this parameter, however, is such that $p$ cannot be negative, then the (non-censored) interval would be $[0,p]$.
We use right-censored intervals, $[p, +\infty)$, when the parameter was set too low, following the same logic.
Figure~\ref{fig:intreg} shows the intervals from the worked example in Section~\ref{sec:workedexample}, where left-censored intervals are depicted as red arrows and right-censored intervals as blue arrows.

In interval regression, we let $y = \beta\mathbf{X} + \epsilon$, where $y$ is a continuous response variable and errors are assumed to be Gaussian, i.e.~$\epsilon \sim \mathcal{N}(0, \sigma^2)$, where $\sigma$ is the standard deviation. In the general case, the log likelihood is composed of four terms:

\begin{equation*}
\begin{split}
    ln L = & - \frac{1}{2} \sum_{j \in \mathcal{C}}{\left\{ \left(\frac{y_j - \mathbf{X}\beta}{\sigma} \right)^2 + log(2\pi\sigma^2) \right\}} \\
           & + \sum_{j \in \mathcal{L}} log\; \Phi \left( \frac{y_{\mathcal{L}_j} - \mathbf{X}\beta}{\sigma} \right) \\
           & + \sum_{j \in \mathcal{R}} log \left\{ 1 - \Phi \left( \frac{y_{\mathcal{R}_j} - \mathbf{X}\beta}{\sigma} \right) \right\} \\
           & + \sum_{j \in \mathcal{I}} log \left\{ \Phi \left( \frac{y_{2_j} - \mathbf{X}\beta}{\sigma} \right) - \Phi \left( \frac{y_{1_j} - \mathbf{X}\beta}{\sigma} \right) \right\},
\end{split}
\end{equation*}
where $\Phi(\cdot)$ is the cumulative normal function.
Each term considers one of the four types of observation: point observations ($\mathcal{C}$), left-censored intervals ($\mathcal{L}$), right-censored intervals ($\mathcal{R}$) and intervals with two end-points ($\mathcal{I}$), respectively \cite{amemiya1973regression}. We note that, in our case, $\mathcal{C}$ is empty as we have no point observations, but a task could allow participants to respond that $p$ is optimal. In our experience, however, participants do not insist that parameterizations are optimal due to a lack of experience with the system under investigation.
Throughout this article, we used the R Survival package to fit interval regression models \cite{therneau2014package}. 

\subsection{Worked Example} 
\label{sec:workedexample}

We present a toy cognitive estimation task to demonstrate the simplest concrete example of collective criticism.

\subsubsection{Objective:}

We placed 568 jelly beans in a jar and used collective criticism to estimate the quantity of jelly beans.
%
%
We determined empirically that the jar could hold $\mathord{\sim}$1000 jelly beans, making the effective range from 0--1000, inclusive.
We used ``greater than'' and ``less than'' as summary retrospective anchors. 
%

\subsubsection{Task:}

For each participant, a number $x$ was sampled uniformly at random from 0-1000, inclusive.
Participants were handed the jar of jelly beans and asked the following question:

\begin{displayquote} \em
    How many jelly beans are in the jar: {\bf \em greater than x} or {\bf \em less than x}?

\end{displayquote}

As follow-up questions, we asked participants to freely estimate the number of jelly beans in the jar, 
and whether they felt critiquing or freely estimating the same quantity was more cognitively demanding.

\subsubsection{Participants:}

We recruited 60 participants during the coffee breaks of a conference organized in our department (27 female, 33 male). 
The participants ranged from PhD students to full professors who had a background in theoretical computer science, optimization or a related field. 

\subsubsection{Results:}

The 60 participants gave 60 critiques and 60 estimates of the number of jelly beans. No data points were excluded from analysis. 
The difference between participants' mean estimate of the number of jelly beans (M = 465.52, 95\% CI [409.98, 521.05])
and the answer derived from collective criticism (M = 454.96, 95\% CI [382.97, 526.95])
was not statistically significant (t(62.56) = -0.260, p = 0.796), 
showing that our approach is as accurate as allowing users to directly estimate.
%
%
Furthermore, a majority of participants (47/60, $p = 1.22 \times 10^{-5}$, binomial test) reported that they perceived critiquing to be less cognitively demanding than free estimation.


\begin{figure*}
    \centering
    \includegraphics[width=\textwidth]{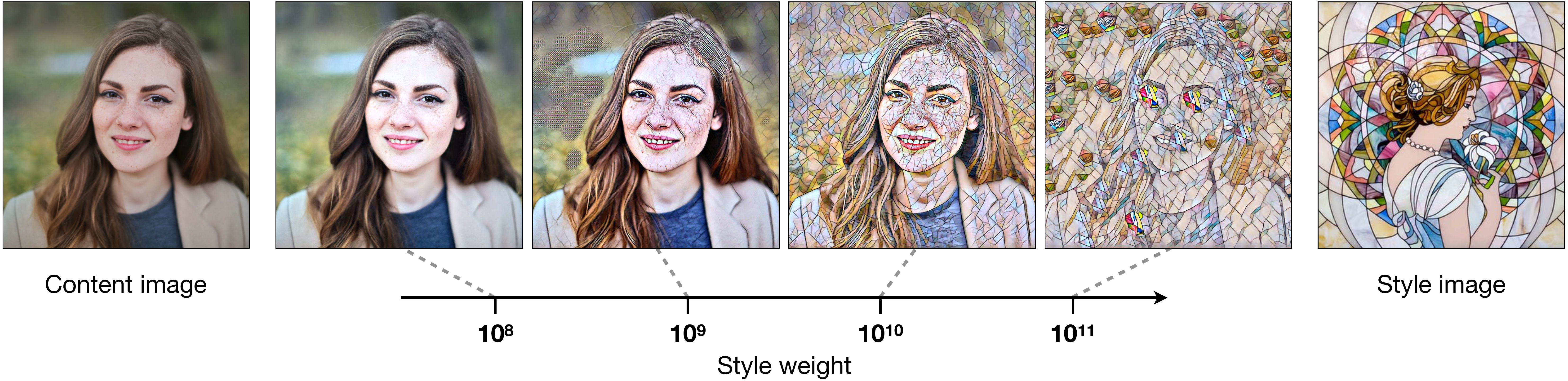}
    \caption{Neural style transfer combines a content image (far left) with a style image (far right). 
    The degree of stylization is controlled by the style weight parameter (middle) which can vary from 
    barely noticeable ($10^8$) to unrecognizable ($10^{11}$).}
    \label{fig:nst}
\end{figure*}

\section{Study 1: Aesthetic Preferences in Neural Style Transfer}
\label{sec:neuraltransfer}


In our first study, we used collective criticism to model users' aesthetic preferences for images generated using neural style transfer \cite{gatys2015neural}.
Neural style transfer combines the content from one image (the content image) and the style from another image, usually an artwork (the style image), 
see Figure~\ref{fig:nst} for an example.
We demonstrate how collective criticism can be used to elicit preferences, model different hypotheses and make practical parameterization decisions.

\subsection{Objective}

Neural style transfer has two hyperparameters: a content weight and a style weight, 
however, if one parameter is kept constant, there is only one free parameter. 
We wanted to identify the highest style weight that could be applied to a photo without the subject becoming unrecognizable.
Furthermore, we hypothesized that different style weights would be optimal for different kinds of photo, 
such as head, waist-up and full body shots, i.e.~we hypothesized that there is an interaction between style weight and photo type.

\subsection{Neural Style Transfer}

\subsubsection{Implementation:} We used the fast neural style transfer implementation included in the PyTorch library\footnote{\url{https://github.com/pytorch/examples/tree/master/fast_neural_style}} 
that is based on perceptual loss \cite{johnson2016perceptual} and instance normalization \cite{ulyanov2016instance}.

\subsubsection{Model Training:} We determined by trial and error that, if the content weight is kept constant at $10^5$, the effective range for style weight is $10^{8}$-$10^{11}$, where higher values result in an output image more heavily influenced by the style image (see Figure~\ref{fig:nst}).
We trained 101 neural style transfer models using the COCO 2014 data set \cite{lin2014microsoft} and used an image of a mosaic included with PyTorch as the style image (see Figure~\ref{fig:nst}, far right). 
Each model had a different style weight parameter where the exponent was incremented by 0.03, i.e.~8.0, 8.03, \dots 10.97, 11.0. 
This increment was chosen empirically so the difference between images generated by consecutive models was imperceptible. Each model was trained for 2 epochs.
%

\subsubsection{Content Images:} We selected photos from a collection of permissively licensed stock photographs\footnote{\url{https://www.pexels.com/license/}} to use as content images.
We selected three categories of portrait: head shots, waist-up and full body shots. 
We identified 39 images of similar size, with 13 photos in each of the three categories. 
All categories included men and women of approximate working age from different ethnicities.

\subsection{Task}

Participants were told we were creating a new website for our research group and wanted to make our photos look more interesting using neural style transfer. We briefly explained the concept of neural style transfer using example images to illustrate different levels of stylization and stated that we wanted output images to be as strongly influenced by the style image as possible without the identity of the person in the content image becoming unrecognizable. 
Each participant was shown 10 randomly sampled images, stylized with randomly sampled style weights. After being shown each image, participants were asked the following question:

\begin{displayquote} \em

    Do you think the image should look {\bf \em more realistic} or {\bf \em more artistic}?
\end{displayquote}


During the study, we logged the photos shown including the photo category, style weights and user critiques. 
Each experiment lasted a total of $\mathord{\sim}$3 minutes. 


\subsection{Participants}

We recruited 31 participants from the Faculty of Science by walking up to people in the corridors of the Department of Computer Science and the Department of Mathematics and Statistics 
(10 female, 21 male). 
All participants were PhD students or postdoctoral researchers.

\subsection{Results}

The 31 participants examined a total of 310 photos: 95 head shots, 117 waist-up shots and 98 full body shots. 

\subsubsection{Baseline:} 

There are no validated questionnaires for rating image stylization and 
pairwise comparisons between all images using all 
models would require an impractical sample size.
Instead, we used a pre-trained model included in the PyTorch distribution 
as a baseline to understand whether 
the developer's original parameterization was sufficient to solve this problem. 
The pre-trained model used a content weight of $10^{5}$ (the same as our models) and a style weight of $10^{10}$.

\subsubsection{Preference Models:}

We fitted two preference-based models using collective criticism. 
In the first model, each study participant was modeled as a random effect (due to repeated measures) and photo type (head, waist-up or full body) was modeled as a fixed effect:

$$y_{ij} = \beta_0 + \beta_1p_{ij} + u_j,$$
where $y$ are intervals derived from critiques, $p$ is the photo type and $u$ are random intercepts per user, $j$.
The second model was identical to the first, but with the photo type term excluded:

$$y_{ij} = \beta_0 + u_j$$

%
The style weights from the first model ($\mathord{log}_{10}$) for 
head (M = 9.71, 95\% CI [9.46, 9.95]),
waist-up, (M = 9.53, 95\% CI [9.31, 9.75]) and
body shots (M = 9.39, 95\% CI [9.14, 9.63]) 
were very similar to one another with overlapping 95\% confidence intervals.
Indeed, the difference in model fit between the two models
was not statistically significant ($\chi^2$ (1.55, N = 310) = 3.14, $p$ = 0.14).
This suggests that the style weight from the second model (M = 9.55, 95\% CI [9.36, 9.73]) is suitable for all three types of photo, assuming all other conditions, such as photo size and the age range of the subject, remain constant. 
Finally, the style weight parameter used in the pre-trained baseline was $10^{10}$, which was 
outside of all four confidence intervals and, therefore, the difference was statistically significant.

\subsection{Summary}

Given these findings, we argue that there is insufficient evidence for using different style weights for each photo type and that the style weight used to train the neural network should be set to $10^{9.55}$ in order to balance style and recognizability. 
However, if we were to collect more data, then it is likely that the confidence intervals would have been narrower and we could recommend the use of category-specific style weights.
This experiment demonstrates how the preferences of an individual (in this case, the original developer) may not match the aesthetic preferences of the group for a specific problem.
This case study only looked at group preferences, but could be extended with additional explanatory variables from a user profile to predict personalized style weights.

\section{Study 2: Prediction of Challenge Perception in Tetris}
\label{sec:tetrisstudy}


In our second study, we show how collective criticism can be used to model users' perception of the challenge experienced while playing the video game Tetris. This example demonstrates how to create a model that could be used as the basis for personalization in an adaptive interactive system. 

\subsection{Objective}

Tetris is a tile-matching video game that gets progressively more challenging as the speed of the game increases (see Section~\ref{sec:tetris} for a description of gameplay). 
A game of Tetris starts out slow, not presenting the player with any challenge. During the endgame, however, Tetris can become frustratingly fast, making the player anxious (see Figure~\ref{fig:tetris}). 
%
%
We wanted to create a model for an adaptive version of Tetris where the level of challenge is personalized for each player. Namely, we wanted to identify how game speed and other contextual factors could be used to keep players feeling challenged, but not overwhelmed. 
This could be viewed as modeling a kind of flow state \cite{csikszentmihalyi1990flow}: one of many considerations that go into game design \cite{baron2012cognitive} and is actively studied in the development of slot machines \cite{schull2005digital}.

\subsection{Tetris}
\label{sec:tetris}

\subsubsection{Gameplay:} In Tetris, players control falling shapes called tetrominoes to achieve a high score. 
Players can move tetrominoes left and right, increase their speed of descent (called a ``soft drop'') 
or force them to immediately drop to the bottom of the play field (a ``hard drop''). 
When the player completes a line (i.e.~a row of blocks without any gaps), 
it disappears and the player's score increases. 
After clearing a fixed number of lines, the difficulty level is increased and, along with it, the speed of the game.
The game ends when a tetromino overlaps the top of the play field. 

\subsubsection{Implementation: }

\begin{figure}
    \centering
    \includegraphics[width=\columnwidth]{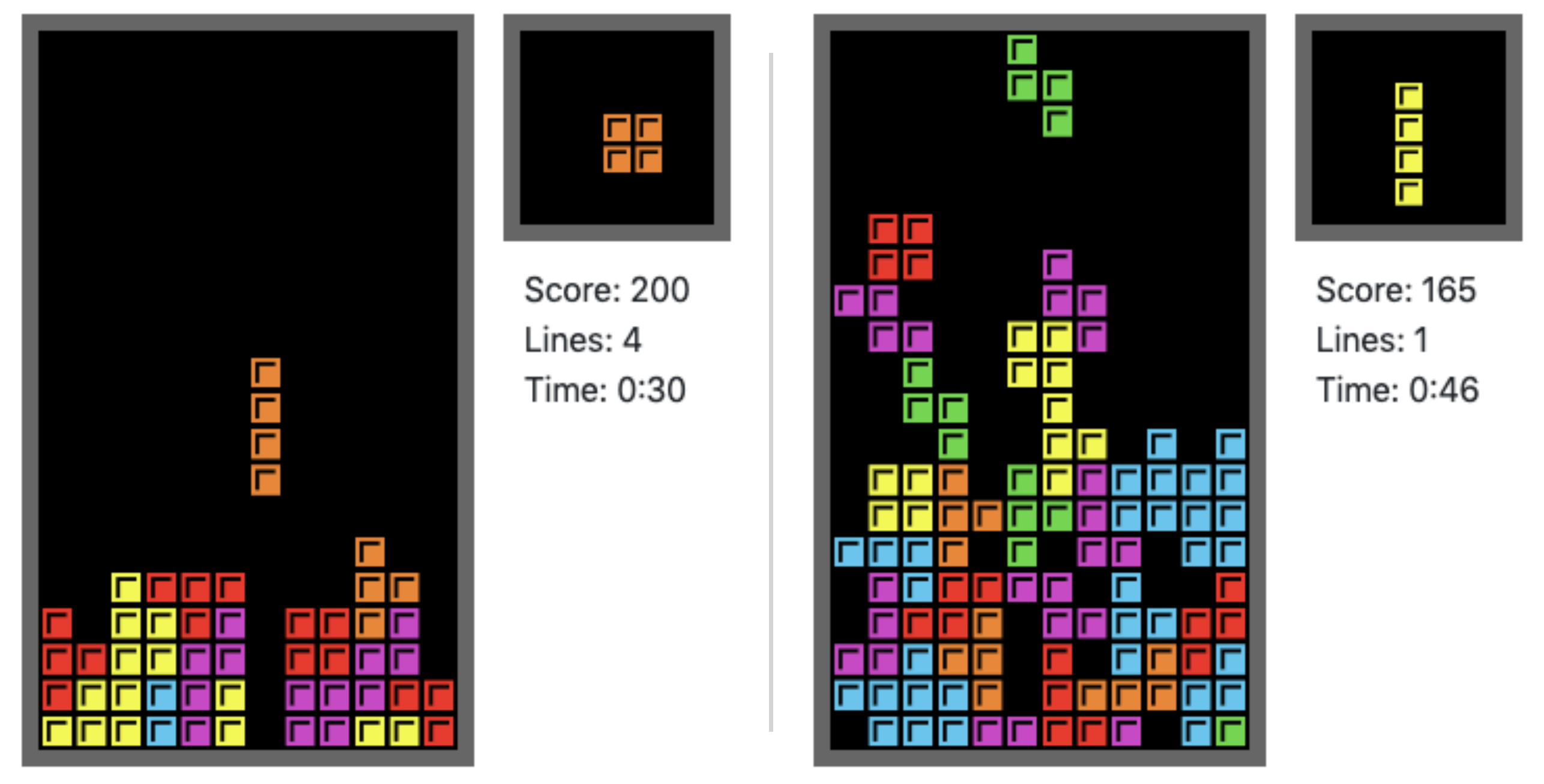}
    \caption{Two screenshots of Tetris: the left image shows the slow (boring) early game, whereas the right image is typical of the faster (stressful) endgame.}
    \label{fig:tetris}
\end{figure}

Figure~\ref{fig:tetris} shows our web-based implementation of Tetris. 
The play field is $12 {\times} 20$ tiles and the surrounding interface shows
the score, the number of lines cleared and a timer indicating how much time is remaining in the task.
In most versions of Tetris, the scoring function is proportional to the level and, therefore, the current game speed, 
i.e.~clearing a line is worth more points at level 2 than level 1. 
In our implementation, however, we used the same scoring function irrespective of the current game speed:
5 points for each placed tetromino and 20 points for each line cleared.
The game speed is determined by the delay in milliseconds for a tetromino to move down the play field by one block. 
We determined by trial and error that the effective range of this delay was 100-600ms, where 100ms gives the fastest speed and 600ms results in the 
slowest speed. 

\subsection{Task}

%
Prior to the task, participants were asked to fill out a background questionnaire to capture 
\begin{enumerate*}[label=(\roman*)]
\item demographic information, 
\item how often they played video games, 
\item their familiarity with Tetris and
\item their opinion of Tetris.
\end{enumerate*}
After completing the questionnaire, participants were allowed to play as many warm-up games of Tetris as they wanted.

During the study, each participant played 3 games of Tetris.
In the first game, the fall delay was sampled uniformly at random from the full effective range, 100-600 inclusive.
In the second and third games, we altered the upper or lower bounds of the delay range to reflect user feedback. 
For example, if in the first round a delay of 300 was too slow, then in the second round the delay would be sampled from the range 100-300 (lower delays mean higher speeds).
Each game of Tetris lasted a maximum of 2 minutes. 
After 2 minutes had expired or the game was lost, participants were asked the following question:

\begin{displayquote} \em
We are collecting data to create a game of Tetris that helps players improve their skill level. 
It should be just fast enough to feel like a challenge.
For a player of your skill level, do you think the game you played should have been {\bf \em faster} or {\bf \em slower}?
\end{displayquote}

During the study, we logged user critiques, game data (fall delay, time spent playing, score, number of lines cleared) 
and interaction data (keyboard events and their associated timestamps).
Each task took a total of $\mathord{\sim}$10 minutes. 

\subsection{Participants}




We recruited 50 participants who were studying at the university library (24 female, 26 male). 
Participants were aged between 20-49 with a median age of 28. 
According to the background questionnaire, over 2/3 of participants played video games at least occasionally 
(never (14), occasionally (23), every week (9), every day (4)), and 
a majority of participants, 45/50, had at least some prior experience of playing Tetris (never (5), a few times (31), many times (12), experienced (2)).
Overall, participants were neutral in their opinion of Tetris (mean = 2.98 on a 5 point scale where 1 = hate and 5 = love).
One participant stated that they hated Tetris, despite having never played the game.

\begin{figure}
    \centering
    \includegraphics[width=\columnwidth]{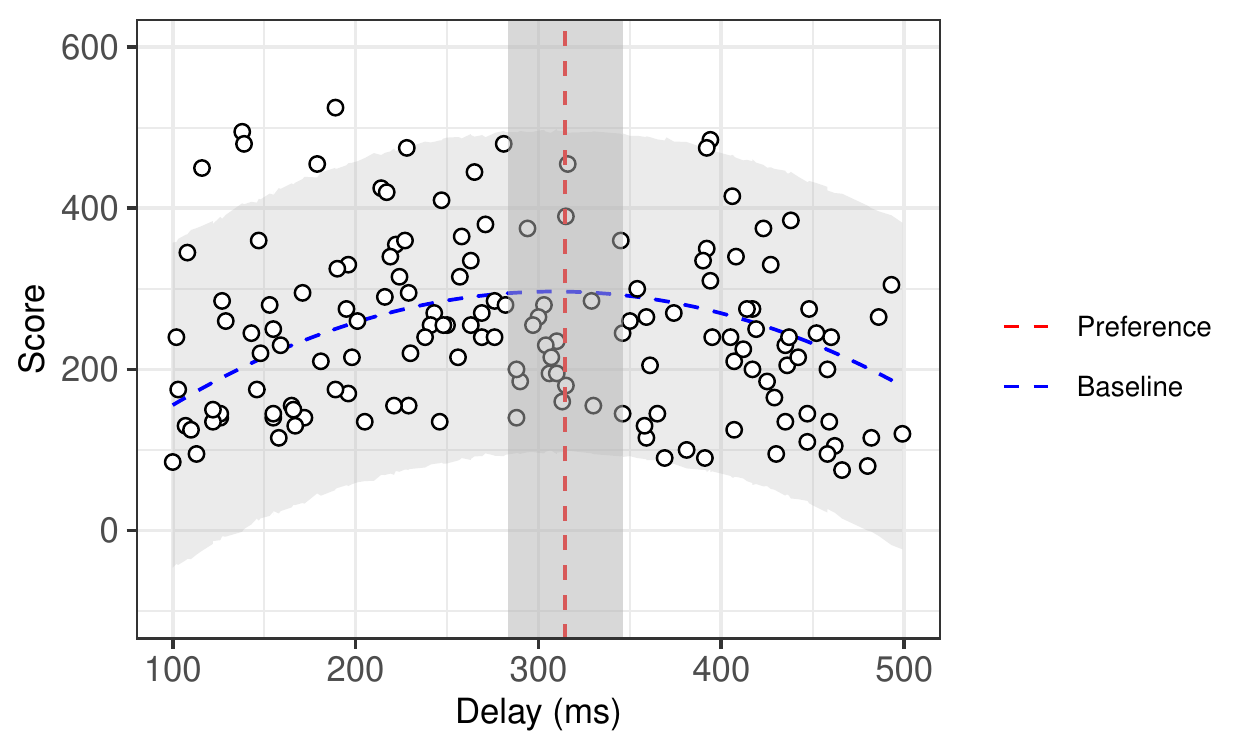}
    \caption{Using player preferences of tetromino fall delay comes to similar conclusions as a 
    quadratic model based on scores. Gray shaded region is the 95\% confidence interval.}
    \label{fig:tetris_quad}
\end{figure}

\subsection{Results}

The 50 participants played a total of 150 games of Tetris. We compared a baseline model that was fit using behavioral data with two models created using collective criticism.

\subsubsection{Baseline:}

We assumed that the fall delay that maximizes the average score would also maximize players' perception that the game was just challenging enough (we are really maximizing the rate of scoring as each experiment has a fixed duration). 
We fitted a linear mixed model with fall delay as linear and quadratic fixed effects and participant as a random effect:
$$y_{ij} = \beta_0 + \beta_1d_{ij}^2 + \beta_2d_{ij} + u_j,$$
where $y$ is the score, $d$ is delay and $u$ are random intercepts per user, $j$.
%
The baseline found that the average score was maximized when the delay was 308.68 ms (see Figure~\ref{fig:tetris_quad}, dashed blue curve).
The baseline does not provide any uncertainty estimates for the optimal delay, 
only standard errors for the coefficients of the delay terms in the model. 
Furthermore, this analysis is based on assumptions and we do not know whether this delay is truly inline with user preferences.

\subsubsection{Static Preference Model:}

We used collective criticism to fit a model for users' speed preferences with no explanatory variables other than participant as a random effect:
$$y_{ij} = \beta_0 + u_j,$$ 
where $y$ are intervals derived from critiques and $u$ are random intercepts per user, $j$.
The preference model found that the optimal delay is 314.70 ms (95\% CI [283.34, 346.05]). This is very similar to 
the point estimate from the baseline model, despite using different response variables and very different assumptions.
Unlike the baseline, this model allows us to estimate the uncertainty in the mean delay (see Figure~\ref{fig:tetris_quad}, dashed red line). As the baseline result (308.68 ms) falls into the 95\% confidence interval for the preference model, there is no statistically significant difference between the two results. However, the preference model has the advantage of having evidence that the optimal delay is inline with user preferences.

\begin{figure}
    \centering
    \includegraphics[width=\columnwidth]{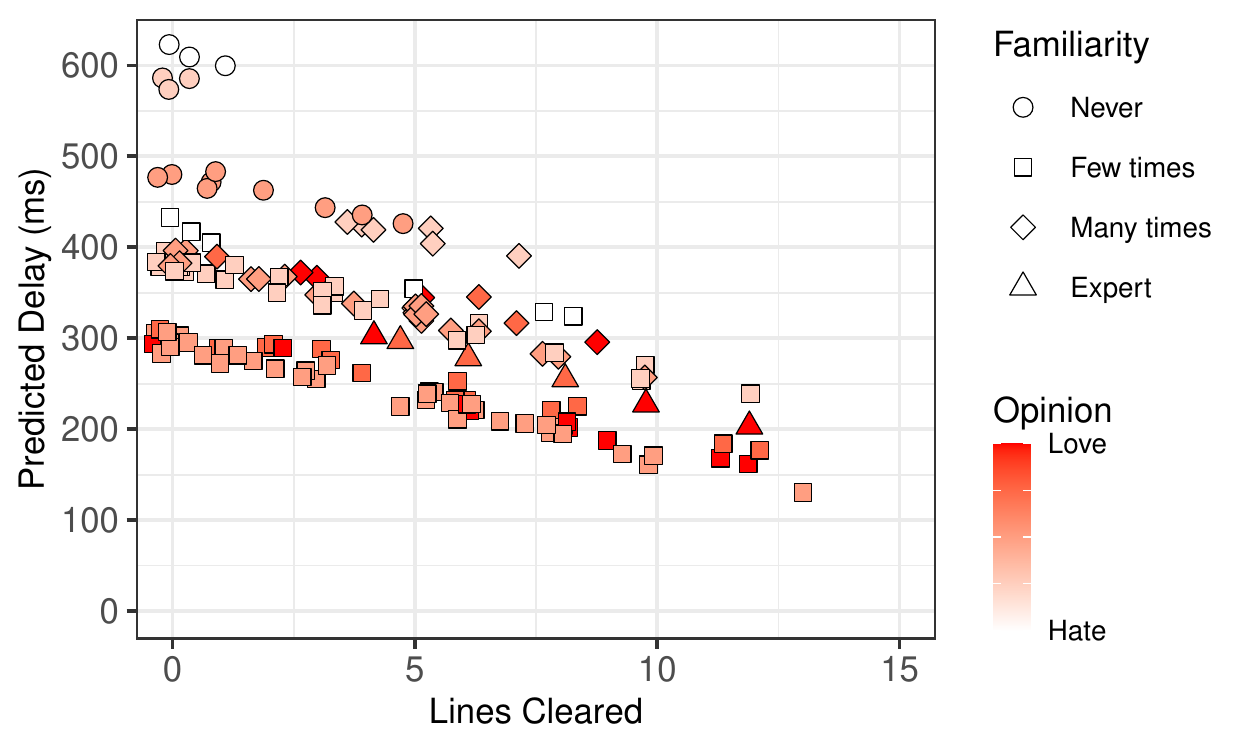}
    \caption{Predicted delays from adaptive preference model. 
    Higher number of lines cleared, Tetris opinion and familiarity predict lower delays are necessary to provide a challenge.}
    \label{fig:tetris_repair}
\end{figure}

\subsubsection{Adaptive Preference Model:}

Predicting the optimal delay for the average participant masks the variability between players and 
we assume novice and expert players will have different ideas of what constitutes a challenge. 
Unfortunately, this is not possible to model with the baseline because the score--a key measure of player ability--is already being used as the response variable. 
However, the preference model can incorporate the score as an additional explanatory variable. 
We tried numerous different models and assessed model fit using AIC.

Figure~\ref{fig:tetris_repair} shows in-sample predictions from the best model we found based on AIC score (AIC = 163.65). 
This model used Tetris familiarity and Tetris opinion from the background questionnaire, as well as the number of lines cleared during a game as explanatory variables (using the game score had a slightly higher AIC, but the two variables were strongly correlated):
$$y_j = \beta_0 + \beta_1s_j + \beta_2f_j + \beta_3o_j + u_j,$$
where $y$ are intervals derived from critiques, $s$ is the number of lines cleared, $f$ is a factor for familiarity, $o$ is a factor for opinion and $u$ are random intercepts per user, $j$.
Figure~\ref{fig:tetris_repair} shows that as player performance (number of lines cleared) increases, 
the predicted fall delay decreases to offer a greater challenge. Similarly, higher Tetris opinions and familiarity tended to correlate with lower predicted fall delay. 
We investigated the use of many additional variables in the model, e.g.~average number of hard drops per tetromino and average time between key presses were highly predictive, however, including them in a model together with number of lines cleared resulted in higher AIC.

\subsection{Summary}

The difference between the baseline and the two models created using collective criticism was the incorporation of preference information, making us confident that we are modeling user experience and not potentially misinterpreting behavioral data. 
While the baseline model would only satisfy the average player, the additional flexibility of freeing up the score variable allows us to create a model for personalization based on player performance (as determined by the prior two minutes of playtime). Furthermore, the preference models directly predict the optimal delay parameter allowing it to be easily integrated into the game itself, whereas the baseline required us to maximize a quadratic function to calculate the optimal result.

\section{Discussion and Limitations}

In this paper, we introduced a critiquing-based approach for modeling the relationships between system parameters and subjective preferences called collective criticism. Collective criticism combines randomized tasks with critiques 
that are transformed into intervals and modeled using interval regression. 
We demonstrated collective criticism using two studies: 
\begin{enumerate*}[label=(\roman*)]
    \item aesthetic preferences in neural style transfer, and 
    \item modeling users' perceptions of challenge in Tetris.
\end{enumerate*}
%
%
In neural style transfer, 
we showed that the optimal parameterization was different from a baseline pre-trained model, but found insufficient evidence in the data collected to support there being an interaction between style weight and photo type. 
In Tetris, we demonstrated that modeling subjective preferences using collective criticism allowed us to create more complex models than with behavioral data alone. Furthermore, we could do so without making assumptions about how behavior maps to preferences. 





There are several limitations to our approach. 
First, collective criticism requires an understanding of regression modeling and how to express experimental design through model specification in order to produce meaningful results. Collective criticism is, therefore, very general, but difficult to use compared to other crowdsourcing approaches where the goal is to generate annotations or a ranking. 
Second, in its present form, participants can only give feedback with respect to a single variable. While this is still very general, there may be situations where two or more variables interact to produce a given user experience, e.g. how a video game feels to play might depend on a combination of character speed and projectile speed. We are currently investigating multiple output regression methods to remedy this limitation.
Third, as we stated earlier, we could not test our approach on crowdsourcing platforms, such as Amazon Mechanical Turk, where studies tend to include trick questions to guard against abuse from bots and disinterested participants. Given the subjective nature of the experiences we are aiming to model and the limited responses used in task design, it is unclear how investigators would ensure data integrity other than simply increasing the sample size and assuming a majority of participants are well-intentioned. This too, will be a topic for future work.
Lastly, 
as we can see in Figure~\ref{fig:intreg}, asking users to critique experiences that are either very high or very low is wasteful after a certain amount of evidence has already accumulated. We plan to investigate combining collective criticism with Thompson sampling \cite{chapelle2011empirical}, where the next data point is sampled from the posterior distribution, so the effective parameter range gets narrower as the experiment proceeds and participants will only be asked to critique the areas of greatest uncertainty. Thompson sampling has the disadvantage that you need to specify the model a priori, which could lead to sample inefficiency if the model is unnecessarily complex. For well-defined problems, however, this should not be an issue.



\begin{acks}
    This work has been supported by Helsinki Institute for Information Technology HIIT.
\end{acks}


\balance
\bibliographystyle{ACM-Reference-Format}
\bibliography{sample-authordraft}


\begin{thebibliography}{49}


\ifx \showCODEN    \undefined \def \showCODEN     #1{\unskip}     \fi
\ifx \showDOI      \undefined \def \showDOI       #1{#1}\fi
\ifx \showISBNx    \undefined \def \showISBNx     #1{\unskip}     \fi
\ifx \showISBNxiii \undefined \def \showISBNxiii  #1{\unskip}     \fi
\ifx \showISSN     \undefined \def \showISSN      #1{\unskip}     \fi
\ifx \showLCCN     \undefined \def \showLCCN      #1{\unskip}     \fi
\ifx \shownote     \undefined \def \shownote      #1{#1}          \fi
\ifx \showarticletitle \undefined \def \showarticletitle #1{#1}   \fi
\ifx \showURL      \undefined \def \showURL       {\relax}        \fi
\providecommand\bibfield[2]{#2}
\providecommand\bibinfo[2]{#2}
\providecommand\natexlab[1]{#1}
\providecommand\showeprint[2][]{arXiv:#2}

\bibitem[\protect\citeauthoryear{Alonso, Rose, and Stewart}{Alonso
  et~al\mbox{.}}{2008}]%
        {alonso2008crowdsourcing}
\bibfield{author}{\bibinfo{person}{Omar Alonso}, \bibinfo{person}{Daniel~E
  Rose}, {and} \bibinfo{person}{Benjamin Stewart}.}
  \bibinfo{year}{2008}\natexlab{}.
\newblock \showarticletitle{Crowdsourcing for relevance evaluation}. In
  \bibinfo{booktitle}{\emph{ACM SigIR Forum}}, Vol.~\bibinfo{volume}{42}. ACM
  New York, NY, USA, \bibinfo{pages}{9--15}.
\newblock


\bibitem[\protect\citeauthoryear{Amemiya}{Amemiya}{1973}]%
        {amemiya1973regression}
\bibfield{author}{\bibinfo{person}{Takeshi Amemiya}.}
  \bibinfo{year}{1973}\natexlab{}.
\newblock \showarticletitle{Regression analysis when the dependent variable is
  truncated normal}.
\newblock \bibinfo{journal}{\emph{Econometrica: Journal of the Econometric
  Society}} (\bibinfo{year}{1973}), \bibinfo{pages}{997--1016}.
\newblock


\bibitem[\protect\citeauthoryear{Aroyo, Dixon, Thain, Redfield, and
  Rosen}{Aroyo et~al\mbox{.}}{2019}]%
        {aroyo2019crowdsourcing}
\bibfield{author}{\bibinfo{person}{Lora Aroyo}, \bibinfo{person}{Lucas Dixon},
  \bibinfo{person}{Nithum Thain}, \bibinfo{person}{Olivia Redfield}, {and}
  \bibinfo{person}{Rachel Rosen}.} \bibinfo{year}{2019}\natexlab{}.
\newblock \showarticletitle{Crowdsourcing subjective tasks: the case study of
  understanding toxicity in online discussions}. In
  \bibinfo{booktitle}{\emph{Companion proceedings of the 2019 world wide web
  conference}}. \bibinfo{pages}{1100--1105}.
\newblock


\bibitem[\protect\citeauthoryear{Awad, Dsouza, Kim, Schulz, Henrich, Shariff,
  Bonnefon, and Rahwan}{Awad et~al\mbox{.}}{2018}]%
        {awad2018moral}
\bibfield{author}{\bibinfo{person}{Edmond Awad}, \bibinfo{person}{Sohan
  Dsouza}, \bibinfo{person}{Richard Kim}, \bibinfo{person}{Jonathan Schulz},
  \bibinfo{person}{Joseph Henrich}, \bibinfo{person}{Azim Shariff},
  \bibinfo{person}{Jean-Fran{\c{c}}ois Bonnefon}, {and} \bibinfo{person}{Iyad
  Rahwan}.} \bibinfo{year}{2018}\natexlab{}.
\newblock \showarticletitle{The moral machine experiment}.
\newblock \bibinfo{journal}{\emph{Nature}} \bibinfo{volume}{563},
  \bibinfo{number}{7729} (\bibinfo{year}{2018}), \bibinfo{pages}{59--64}.
\newblock


\bibitem[\protect\citeauthoryear{Azadgoli, Gould, Vartanian, and
  Patel}{Azadgoli et~al\mbox{.}}{2019}]%
        {azadgoli2019public}
\bibfield{author}{\bibinfo{person}{Beina Azadgoli}, \bibinfo{person}{Daniel~J
  Gould}, \bibinfo{person}{Emma Vartanian}, {and} \bibinfo{person}{Ketan~M
  Patel}.} \bibinfo{year}{2019}\natexlab{}.
\newblock \showarticletitle{The public’s perception on breast and nipple
  reconstruction: a crowdsourcing-based assessment}.
\newblock \bibinfo{journal}{\emph{Aesthetic surgery journal}}
  \bibinfo{volume}{39}, \bibinfo{number}{9} (\bibinfo{year}{2019}),
  \bibinfo{pages}{NP370--NP376}.
\newblock


\bibitem[\protect\citeauthoryear{Bakshi, Kaur, Kaur, and Kaur}{Bakshi
  et~al\mbox{.}}{2016}]%
        {bakshi2016opinion}
\bibfield{author}{\bibinfo{person}{Rushlene~Kaur Bakshi},
  \bibinfo{person}{Navneet Kaur}, \bibinfo{person}{Ravneet Kaur}, {and}
  \bibinfo{person}{Gurpreet Kaur}.} \bibinfo{year}{2016}\natexlab{}.
\newblock \showarticletitle{Opinion mining and sentiment analysis}. In
  \bibinfo{booktitle}{\emph{2016 3rd international conference on computing for
  sustainable global development (INDIACom)}}. IEEE, \bibinfo{pages}{452--455}.
\newblock


\bibitem[\protect\citeauthoryear{Baron}{Baron}{2012}]%
        {baron2012cognitive}
\bibfield{author}{\bibinfo{person}{Sean Baron}.}
  \bibinfo{year}{2012}\natexlab{}.
\newblock \showarticletitle{Cognitive flow: the psychology of great game
  design}.
\newblock \bibinfo{journal}{\emph{Gamastura}} (\bibinfo{year}{2012}).
\newblock


\bibitem[\protect\citeauthoryear{Behrend, Sharek, Meade, and Wiebe}{Behrend
  et~al\mbox{.}}{2011}]%
        {behrend2011viability}
\bibfield{author}{\bibinfo{person}{Tara~S Behrend}, \bibinfo{person}{David~J
  Sharek}, \bibinfo{person}{Adam~W Meade}, {and} \bibinfo{person}{Eric~N
  Wiebe}.} \bibinfo{year}{2011}\natexlab{}.
\newblock \showarticletitle{The viability of crowdsourcing for survey
  research}.
\newblock \bibinfo{journal}{\emph{Behavior research methods}}
  \bibinfo{volume}{43}, \bibinfo{number}{3} (\bibinfo{year}{2011}),
  \bibinfo{pages}{800--813}.
\newblock


\bibitem[\protect\citeauthoryear{Bernhaupt}{Bernhaupt}{2010}]%
        {bernhaupt2010user}
\bibfield{author}{\bibinfo{person}{Regina Bernhaupt}.}
  \bibinfo{year}{2010}\natexlab{}.
\newblock \showarticletitle{User experience evaluation in entertainment}.
\newblock In \bibinfo{booktitle}{\emph{Evaluating user experience in games}}.
  \bibinfo{publisher}{Springer}, \bibinfo{pages}{3--7}.
\newblock


\bibitem[\protect\citeauthoryear{Burke}{Burke}{2002}]%
        {burke2002interactive}
\bibfield{author}{\bibinfo{person}{Robin Burke}.}
  \bibinfo{year}{2002}\natexlab{}.
\newblock \showarticletitle{Interactive critiquing forcatalog navigation in
  e-commerce}.
\newblock \bibinfo{journal}{\emph{Artificial Intelligence Review}}
  \bibinfo{volume}{18}, \bibinfo{number}{3} (\bibinfo{year}{2002}),
  \bibinfo{pages}{245--267}.
\newblock


\bibitem[\protect\citeauthoryear{Burke, Hammond, and Yound}{Burke
  et~al\mbox{.}}{1997}]%
        {burke1997findme}
\bibfield{author}{\bibinfo{person}{Robin~D Burke}, \bibinfo{person}{Kristian~J
  Hammond}, {and} \bibinfo{person}{BC Yound}.} \bibinfo{year}{1997}\natexlab{}.
\newblock \showarticletitle{The FindMe approach to assisted browsing}.
\newblock \bibinfo{journal}{\emph{IEEE Expert}} \bibinfo{volume}{12},
  \bibinfo{number}{4} (\bibinfo{year}{1997}), \bibinfo{pages}{32--40}.
\newblock


\bibitem[\protect\citeauthoryear{Chapelle and Li}{Chapelle and Li}{2011}]%
        {chapelle2011empirical}
\bibfield{author}{\bibinfo{person}{Olivier Chapelle} {and}
  \bibinfo{person}{Lihong Li}.} \bibinfo{year}{2011}\natexlab{}.
\newblock \showarticletitle{An empirical evaluation of thompson sampling}. In
  \bibinfo{booktitle}{\emph{Advances in neural information processing
  systems}}. \bibinfo{pages}{2249--2257}.
\newblock


\bibitem[\protect\citeauthoryear{Chen and Pu}{Chen and Pu}{2012}]%
        {chen2012critiquing}
\bibfield{author}{\bibinfo{person}{Li Chen} {and} \bibinfo{person}{Pearl Pu}.}
  \bibinfo{year}{2012}\natexlab{}.
\newblock \showarticletitle{Critiquing-based recommenders: survey and emerging
  trends}.
\newblock \bibinfo{journal}{\emph{User Modeling and User-Adapted Interaction}}
  \bibinfo{volume}{22}, \bibinfo{number}{1} (\bibinfo{year}{2012}),
  \bibinfo{pages}{125--150}.
\newblock


\bibitem[\protect\citeauthoryear{Crump, McDonnell, and Gureckis}{Crump
  et~al\mbox{.}}{2013}]%
        {crump2013evaluating}
\bibfield{author}{\bibinfo{person}{Matthew~JC Crump}, \bibinfo{person}{John~V
  McDonnell}, {and} \bibinfo{person}{Todd~M Gureckis}.}
  \bibinfo{year}{2013}\natexlab{}.
\newblock \showarticletitle{Evaluating Amazon's Mechanical Turk as a tool for
  experimental behavioral research}.
\newblock \bibinfo{journal}{\emph{PloS one}} \bibinfo{volume}{8},
  \bibinfo{number}{3} (\bibinfo{year}{2013}), \bibinfo{pages}{e57410}.
\newblock


\bibitem[\protect\citeauthoryear{Csikzentmihaly}{Csikzentmihaly}{1990}]%
        {csikszentmihalyi1990flow}
\bibfield{author}{\bibinfo{person}{Mihaly Csikzentmihaly}.}
  \bibinfo{year}{1990}\natexlab{}.
\newblock \bibinfo{booktitle}{\emph{Flow: The psychology of optimal
  experience}}. Vol.~\bibinfo{volume}{1990}.
\newblock \bibinfo{publisher}{Harper \& Row New York}.
\newblock


\bibitem[\protect\citeauthoryear{Dev, Villar, and Lau}{Dev
  et~al\mbox{.}}{2017}]%
        {dev2017polygons}
\bibfield{author}{\bibinfo{person}{Kapil Dev}, \bibinfo{person}{Nicolas
  Villar}, {and} \bibinfo{person}{Manfred Lau}.}
  \bibinfo{year}{2017}\natexlab{}.
\newblock \showarticletitle{Polygons, points, or voxels? stimuli selection for
  crowdsourcing aesthetics preferences of 3D shape pairs}. In
  \bibinfo{booktitle}{\emph{Proceedings of the symposium on Computational
  Aesthetics}}. \bibinfo{pages}{1--7}.
\newblock


\bibitem[\protect\citeauthoryear{Expressions}{Expressions}{[n.\,d.]}]%
        {expressions33mirror}
\bibfield{author}{\bibinfo{person}{Most~Attractive Expressions}.}
  \bibinfo{year}{[n.\,d.]}\natexlab{}.
\newblock \showarticletitle{Mirror Mirror: Crowdsourcing Better Portraits}.
\newblock \bibinfo{journal}{\emph{To appear in ACM TOG}}  \bibinfo{volume}{33}
  (\bibinfo{year}{[n.\,d.]}), \bibinfo{pages}{6}.
\newblock


\bibitem[\protect\citeauthoryear{Faltings, Pu, Torrens, and Viappiani}{Faltings
  et~al\mbox{.}}{2004}]%
        {faltings2004designing}
\bibfield{author}{\bibinfo{person}{Boi Faltings}, \bibinfo{person}{Pearl Pu},
  \bibinfo{person}{Marc Torrens}, {and} \bibinfo{person}{Paolo Viappiani}.}
  \bibinfo{year}{2004}\natexlab{}.
\newblock \showarticletitle{Designing example-critiquing interaction}. In
  \bibinfo{booktitle}{\emph{Proceedings of the 9th international conference on
  Intelligent user interfaces}}. \bibinfo{pages}{22--29}.
\newblock


\bibitem[\protect\citeauthoryear{Filatova}{Filatova}{2012}]%
        {filatova2012irony}
\bibfield{author}{\bibinfo{person}{Elena Filatova}.}
  \bibinfo{year}{2012}\natexlab{}.
\newblock \showarticletitle{Irony and Sarcasm: Corpus Generation and Analysis
  Using Crowdsourcing.}. In \bibinfo{booktitle}{\emph{Lrec}}. Citeseer,
  \bibinfo{pages}{392--398}.
\newblock


\bibitem[\protect\citeauthoryear{Frojo, Kareh, Probst, Rector, Plikaitis, Lund,
  and Lin}{Frojo et~al\mbox{.}}{2021}]%
        {frojo2021defining}
\bibfield{author}{\bibinfo{person}{Gianfranco Frojo}, \bibinfo{person}{Aurora~M
  Kareh}, \bibinfo{person}{Kenneth~X Probst}, \bibinfo{person}{Jeffrey~D
  Rector}, \bibinfo{person}{Christina~M Plikaitis}, \bibinfo{person}{Herluf~G
  Lund}, {and} \bibinfo{person}{Alexander~Y Lin}.}
  \bibinfo{year}{2021}\natexlab{}.
\newblock \showarticletitle{Defining Ideal External Female Genital Anatomy Via
  Crowdsourcing Analysis}.
\newblock \bibinfo{journal}{\emph{Aesthetic surgery journal}}
  (\bibinfo{year}{2021}).
\newblock


\bibitem[\protect\citeauthoryear{Gatys, Ecker, and Bethge}{Gatys
  et~al\mbox{.}}{2015}]%
        {gatys2015neural}
\bibfield{author}{\bibinfo{person}{Leon~A Gatys}, \bibinfo{person}{Alexander~S
  Ecker}, {and} \bibinfo{person}{Matthias Bethge}.}
  \bibinfo{year}{2015}\natexlab{}.
\newblock \showarticletitle{A neural algorithm of artistic style}.
\newblock \bibinfo{journal}{\emph{arXiv preprint arXiv:1508.06576}}
  (\bibinfo{year}{2015}).
\newblock


\bibitem[\protect\citeauthoryear{Hassenzahl}{Hassenzahl}{2008}]%
        {hassenzahl2008user}
\bibfield{author}{\bibinfo{person}{Marc Hassenzahl}.}
  \bibinfo{year}{2008}\natexlab{}.
\newblock \showarticletitle{User experience (UX) towards an experiential
  perspective on product quality}. In \bibinfo{booktitle}{\emph{Proceedings of
  the 20th Conference on l'Interaction Homme-Machine}}.
  \bibinfo{pages}{11--15}.
\newblock


\bibitem[\protect\citeauthoryear{Hassenzahl and Tractinsky}{Hassenzahl and
  Tractinsky}{2006}]%
        {hassenzahl2006user}
\bibfield{author}{\bibinfo{person}{Marc Hassenzahl} {and} \bibinfo{person}{Noam
  Tractinsky}.} \bibinfo{year}{2006}\natexlab{}.
\newblock \showarticletitle{User experience-a research agenda}.
\newblock \bibinfo{journal}{\emph{Behaviour \& information technology}}
  \bibinfo{volume}{25}, \bibinfo{number}{2} (\bibinfo{year}{2006}),
  \bibinfo{pages}{91--97}.
\newblock


\bibitem[\protect\citeauthoryear{Jannach, Manzoor, Cai, and Chen}{Jannach
  et~al\mbox{.}}{2020}]%
        {jannach2020survey}
\bibfield{author}{\bibinfo{person}{Dietmar Jannach}, \bibinfo{person}{Ahtsham
  Manzoor}, \bibinfo{person}{Wanling Cai}, {and} \bibinfo{person}{Li Chen}.}
  \bibinfo{year}{2020}\natexlab{}.
\newblock \showarticletitle{A survey on conversational recommender systems}.
\newblock \bibinfo{journal}{\emph{arXiv preprint arXiv:2004.00646}}
  (\bibinfo{year}{2020}).
\newblock


\bibitem[\protect\citeauthoryear{Jin, Cai, Chen, Htun, and Verbert}{Jin
  et~al\mbox{.}}{2019}]%
        {jin2019musicbot}
\bibfield{author}{\bibinfo{person}{Yucheng Jin}, \bibinfo{person}{Wanling Cai},
  \bibinfo{person}{Li Chen}, \bibinfo{person}{Nyi~Nyi Htun}, {and}
  \bibinfo{person}{Katrien Verbert}.} \bibinfo{year}{2019}\natexlab{}.
\newblock \showarticletitle{MusicBot: Evaluating critiquing-based music
  recommenders with conversational interaction}. In
  \bibinfo{booktitle}{\emph{Proceedings of the 28th ACM International
  Conference on Information and Knowledge Management}}.
  \bibinfo{pages}{951--960}.
\newblock


\bibitem[\protect\citeauthoryear{Jin, Carman, Zhu, and Xiang}{Jin
  et~al\mbox{.}}{2020}]%
        {jin2020technical}
\bibfield{author}{\bibinfo{person}{Yuan Jin}, \bibinfo{person}{Mark Carman},
  \bibinfo{person}{Ye Zhu}, {and} \bibinfo{person}{Yong Xiang}.}
  \bibinfo{year}{2020}\natexlab{}.
\newblock \showarticletitle{A technical survey on statistical modelling and
  design methods for crowdsourcing quality control}.
\newblock \bibinfo{journal}{\emph{Artificial Intelligence}}
  (\bibinfo{year}{2020}), \bibinfo{pages}{103351}.
\newblock


\bibitem[\protect\citeauthoryear{Johnson, Alahi, and Fei-Fei}{Johnson
  et~al\mbox{.}}{2016}]%
        {johnson2016perceptual}
\bibfield{author}{\bibinfo{person}{Justin Johnson}, \bibinfo{person}{Alexandre
  Alahi}, {and} \bibinfo{person}{Li Fei-Fei}.} \bibinfo{year}{2016}\natexlab{}.
\newblock \showarticletitle{Perceptual losses for real-time style transfer and
  super-resolution}. In \bibinfo{booktitle}{\emph{European conference on
  computer vision}}. Springer, \bibinfo{pages}{694--711}.
\newblock


\bibitem[\protect\citeauthoryear{Kohavi and Longbotham}{Kohavi and
  Longbotham}{2017}]%
        {kohavi2017online}
\bibfield{author}{\bibinfo{person}{Ron Kohavi} {and} \bibinfo{person}{Roger
  Longbotham}.} \bibinfo{year}{2017}\natexlab{}.
\newblock \showarticletitle{Online Controlled Experiments and A/B Testing.}
\newblock \bibinfo{journal}{\emph{Encyclopedia of machine learning and data
  mining}} \bibinfo{volume}{7}, \bibinfo{number}{8} (\bibinfo{year}{2017}),
  \bibinfo{pages}{922--929}.
\newblock


\bibitem[\protect\citeauthoryear{Kohavi and Thomke}{Kohavi and Thomke}{2017}]%
        {kohavi2017surprising}
\bibfield{author}{\bibinfo{person}{Ron Kohavi} {and} \bibinfo{person}{Stefan
  Thomke}.} \bibinfo{year}{2017}\natexlab{}.
\newblock \showarticletitle{The surprising power of online experiments}.
\newblock \bibinfo{journal}{\emph{Harvard Business Review}}
  (\bibinfo{year}{2017}).
\newblock


\bibitem[\protect\citeauthoryear{Lin, Maire, Belongie, Hays, Perona, Ramanan,
  Doll{\'a}r, and Zitnick}{Lin et~al\mbox{.}}{2014}]%
        {lin2014microsoft}
\bibfield{author}{\bibinfo{person}{Tsung-Yi Lin}, \bibinfo{person}{Michael
  Maire}, \bibinfo{person}{Serge Belongie}, \bibinfo{person}{James Hays},
  \bibinfo{person}{Pietro Perona}, \bibinfo{person}{Deva Ramanan},
  \bibinfo{person}{Piotr Doll{\'a}r}, {and} \bibinfo{person}{C~Lawrence
  Zitnick}.} \bibinfo{year}{2014}\natexlab{}.
\newblock \showarticletitle{Microsoft coco: Common objects in context}. In
  \bibinfo{booktitle}{\emph{European conference on computer vision}}. Springer,
  \bibinfo{pages}{740--755}.
\newblock


\bibitem[\protect\citeauthoryear{Lindgaard and Kirakowski}{Lindgaard and
  Kirakowski}{2013}]%
        {lindgaard2013introduction}
\bibfield{author}{\bibinfo{person}{Gitte Lindgaard} {and}
  \bibinfo{person}{Jurek Kirakowski}.} \bibinfo{year}{2013}\natexlab{}.
\newblock \showarticletitle{The tricky landscape of developing rating scales in
  HCI}.
\newblock \bibinfo{journal}{\emph{Interacting with Computers}}
  \bibinfo{volume}{25}, \bibinfo{number}{4} (\bibinfo{year}{2013}),
  \bibinfo{pages}{271--277}.
\newblock


\bibitem[\protect\citeauthoryear{Massie, Sood, Nolan, Sasson, Swanson,
  Morrison, and Placik}{Massie et~al\mbox{.}}{2021}]%
        {massie2021defining}
\bibfield{author}{\bibinfo{person}{Jonathan~P Massie}, \bibinfo{person}{Rachita
  Sood}, \bibinfo{person}{Ian~T Nolan}, \bibinfo{person}{Daniel~C Sasson},
  \bibinfo{person}{Marco Swanson}, \bibinfo{person}{Shane~D Morrison}, {and}
  \bibinfo{person}{Otto Placik}.} \bibinfo{year}{2021}\natexlab{}.
\newblock \showarticletitle{Defining aesthetic preferences for the penis: a
  photogrammetric and crowdsourcing analysis}.
\newblock \bibinfo{journal}{\emph{Aesthetic surgery journal}}
  (\bibinfo{year}{2021}).
\newblock


\bibitem[\protect\citeauthoryear{Medlar, Pyykk{\"o}, and Glowacka}{Medlar
  et~al\mbox{.}}{2017}]%
        {medlar2017towards}
\bibfield{author}{\bibinfo{person}{Alan Medlar}, \bibinfo{person}{Joel
  Pyykk{\"o}}, {and} \bibinfo{person}{Dorota Glowacka}.}
  \bibinfo{year}{2017}\natexlab{}.
\newblock \showarticletitle{Towards fine-grained adaptation of
  exploration/exploitation in information retrieval}. In
  \bibinfo{booktitle}{\emph{Proceedings of the 22nd International Conference on
  Intelligent User Interfaces}}. \bibinfo{pages}{623--627}.
\newblock


\bibitem[\protect\citeauthoryear{Perez-Ortiz and Mantiuk}{Perez-Ortiz and
  Mantiuk}{2017}]%
        {perez2017practical}
\bibfield{author}{\bibinfo{person}{Maria Perez-Ortiz} {and}
  \bibinfo{person}{Rafal~K Mantiuk}.} \bibinfo{year}{2017}\natexlab{}.
\newblock \showarticletitle{A practical guide and software for analysing
  pairwise comparison experiments}.
\newblock \bibinfo{journal}{\emph{arXiv preprint arXiv:1712.03686}}
  (\bibinfo{year}{2017}).
\newblock


\bibitem[\protect\citeauthoryear{Pu and Faltings}{Pu and Faltings}{2004}]%
        {pu2004decision}
\bibfield{author}{\bibinfo{person}{Pearl Pu} {and} \bibinfo{person}{Boi
  Faltings}.} \bibinfo{year}{2004}\natexlab{}.
\newblock \showarticletitle{Decision tradeoff using example-critiquing and
  constraint programming}.
\newblock \bibinfo{journal}{\emph{Constraints}} \bibinfo{volume}{9},
  \bibinfo{number}{4} (\bibinfo{year}{2004}), \bibinfo{pages}{289--310}.
\newblock


\bibitem[\protect\citeauthoryear{S{\'a}nchez, Vela, Simarro, and
  Padilla-Zea}{S{\'a}nchez et~al\mbox{.}}{2012}]%
        {sanchez2012playability}
\bibfield{author}{\bibinfo{person}{Jos{\'e} Luis~Gonz{\'a}lez S{\'a}nchez},
  \bibinfo{person}{Francisco Luis~Guti{\'e}rrez Vela},
  \bibinfo{person}{Francisco~Montero Simarro}, {and} \bibinfo{person}{Natalia
  Padilla-Zea}.} \bibinfo{year}{2012}\natexlab{}.
\newblock \showarticletitle{Playability: analysing user experience in video
  games}.
\newblock \bibinfo{journal}{\emph{Behaviour \& Information Technology}}
  \bibinfo{volume}{31}, \bibinfo{number}{10} (\bibinfo{year}{2012}),
  \bibinfo{pages}{1033--1054}.
\newblock


\bibitem[\protect\citeauthoryear{Schull}{Schull}{2005}]%
        {schull2005digital}
\bibfield{author}{\bibinfo{person}{Natasha~Dow Schull}.}
  \bibinfo{year}{2005}\natexlab{}.
\newblock \showarticletitle{Digital gambling: The coincidence of desire and
  design}.
\newblock \bibinfo{journal}{\emph{The Annals of the American Academy of
  Political and Social Science}} \bibinfo{volume}{597}, \bibinfo{number}{1}
  (\bibinfo{year}{2005}), \bibinfo{pages}{65--81}.
\newblock


\bibitem[\protect\citeauthoryear{Shaker, Yannakakis, and Togelius}{Shaker
  et~al\mbox{.}}{2012}]%
        {shaker2012crowdsourcing}
\bibfield{author}{\bibinfo{person}{Noor Shaker}, \bibinfo{person}{Georgios~N
  Yannakakis}, {and} \bibinfo{person}{Julian Togelius}.}
  \bibinfo{year}{2012}\natexlab{}.
\newblock \showarticletitle{Crowdsourcing the aesthetics of platform games}.
\newblock \bibinfo{journal}{\emph{IEEE Transactions on Computational
  Intelligence and AI in Games}} \bibinfo{volume}{5}, \bibinfo{number}{3}
  (\bibinfo{year}{2012}), \bibinfo{pages}{276--290}.
\newblock


\bibitem[\protect\citeauthoryear{Suchyta, Azad, Patel, Khosla, Lorenz, and
  Nazerali}{Suchyta et~al\mbox{.}}{2020}]%
        {suchyta2020applied}
\bibfield{author}{\bibinfo{person}{Marissa Suchyta}, \bibinfo{person}{Amee
  Azad}, \bibinfo{person}{Ashraf~A Patel}, \bibinfo{person}{Rohit~K Khosla},
  \bibinfo{person}{H~Peter Lorenz}, {and} \bibinfo{person}{Rahim~S Nazerali}.}
  \bibinfo{year}{2020}\natexlab{}.
\newblock \showarticletitle{Applied online crowdsourcing in plastic and
  reconstructive surgery: a comparison of aesthetic outcomes in unilateral
  cleft lip repair techniques}.
\newblock \bibinfo{journal}{\emph{Annals of plastic surgery}}
  \bibinfo{volume}{84}, \bibinfo{number}{5S} (\bibinfo{year}{2020}),
  \bibinfo{pages}{S307--S310}.
\newblock


\bibitem[\protect\citeauthoryear{Swink}{Swink}{2008}]%
        {swink2008game}
\bibfield{author}{\bibinfo{person}{Steve Swink}.}
  \bibinfo{year}{2008}\natexlab{}.
\newblock \bibinfo{booktitle}{\emph{Game feel: a game designer's guide to
  virtual sensation}}.
\newblock \bibinfo{publisher}{CRC Press}.
\newblock


\bibitem[\protect\citeauthoryear{Therneau and Lumley}{Therneau and
  Lumley}{2014}]%
        {therneau2014package}
\bibfield{author}{\bibinfo{person}{Terry~M Therneau} {and}
  \bibinfo{person}{Thomas Lumley}.} \bibinfo{year}{2014}\natexlab{}.
\newblock \showarticletitle{Package ‘survival’}.
\newblock \bibinfo{journal}{\emph{Survival analysis Published on CRAN}}
  \bibinfo{volume}{2} (\bibinfo{year}{2014}), \bibinfo{pages}{3}.
\newblock


\bibitem[\protect\citeauthoryear{Tse, Oh, Gruss, Hopper, and Birgfeld}{Tse
  et~al\mbox{.}}{2016}]%
        {tse2016crowdsourcing}
\bibfield{author}{\bibinfo{person}{Raymond~W Tse}, \bibinfo{person}{Eugene Oh},
  \bibinfo{person}{Joseph~S Gruss}, \bibinfo{person}{Richard~A Hopper}, {and}
  \bibinfo{person}{Craig~B Birgfeld}.} \bibinfo{year}{2016}\natexlab{}.
\newblock \showarticletitle{Crowdsourcing as a novel method to evaluate
  aesthetic outcomes of treatment for unilateral cleft lip}.
\newblock \bibinfo{journal}{\emph{Plastic and reconstructive surgery}}
  \bibinfo{volume}{138}, \bibinfo{number}{4} (\bibinfo{year}{2016}),
  \bibinfo{pages}{864--874}.
\newblock


\bibitem[\protect\citeauthoryear{Tufekci}{Tufekci}{2018}]%
        {tufekci2018youtube}
\bibfield{author}{\bibinfo{person}{Zeynep Tufekci}.}
  \bibinfo{year}{2018}\natexlab{}.
\newblock \showarticletitle{{YouTube}, the great radicalizer}.
\newblock \bibinfo{journal}{\emph{The New York Times}}  \bibinfo{volume}{10}
  (\bibinfo{year}{2018}), \bibinfo{pages}{2018}.
\newblock


\bibitem[\protect\citeauthoryear{Ulyanov, Vedaldi, and Lempitsky}{Ulyanov
  et~al\mbox{.}}{2016}]%
        {ulyanov2016instance}
\bibfield{author}{\bibinfo{person}{Dmitry Ulyanov}, \bibinfo{person}{Andrea
  Vedaldi}, {and} \bibinfo{person}{Victor Lempitsky}.}
  \bibinfo{year}{2016}\natexlab{}.
\newblock \showarticletitle{Instance normalization: The missing ingredient for
  fast stylization}.
\newblock \bibinfo{journal}{\emph{arXiv preprint arXiv:1607.08022}}
  (\bibinfo{year}{2016}).
\newblock


\bibitem[\protect\citeauthoryear{Van~Berkel, Goncalves, Hettiachchi,
  Wijenayake, Kelly, and Kostakos}{Van~Berkel et~al\mbox{.}}{2019}]%
        {van2019crowdsourcing}
\bibfield{author}{\bibinfo{person}{Niels Van~Berkel}, \bibinfo{person}{Jorge
  Goncalves}, \bibinfo{person}{Danula Hettiachchi}, \bibinfo{person}{Senuri
  Wijenayake}, \bibinfo{person}{Ryan~M Kelly}, {and} \bibinfo{person}{Vassilis
  Kostakos}.} \bibinfo{year}{2019}\natexlab{}.
\newblock \showarticletitle{Crowdsourcing perceptions of fair predictors for
  machine learning: a recidivism case study}.
\newblock \bibinfo{journal}{\emph{Proceedings of the ACM on Human-Computer
  Interaction}} \bibinfo{volume}{3}, \bibinfo{number}{CSCW}
  (\bibinfo{year}{2019}), \bibinfo{pages}{1--21}.
\newblock


\bibitem[\protect\citeauthoryear{Vartanian, Gould, Hammoudeh, Azadgoli,
  Stevens, and Macias}{Vartanian et~al\mbox{.}}{2018}]%
        {vartanian2018ideal}
\bibfield{author}{\bibinfo{person}{Emma Vartanian}, \bibinfo{person}{Daniel~J
  Gould}, \bibinfo{person}{Ziyad~S Hammoudeh}, \bibinfo{person}{Beina
  Azadgoli}, \bibinfo{person}{W~Grant Stevens}, {and} \bibinfo{person}{Luis~H
  Macias}.} \bibinfo{year}{2018}\natexlab{}.
\newblock \showarticletitle{The ideal thigh: a crowdsourcing-based assessment
  of ideal thigh aesthetic and implications for gluteal fat grafting}.
\newblock \bibinfo{journal}{\emph{Aesthetic surgery journal}}
  \bibinfo{volume}{38}, \bibinfo{number}{8} (\bibinfo{year}{2018}),
  \bibinfo{pages}{861--869}.
\newblock


\bibitem[\protect\citeauthoryear{Viappiani, Faltings, and Pu}{Viappiani
  et~al\mbox{.}}{2006}]%
        {viappiani2006preference}
\bibfield{author}{\bibinfo{person}{Paolo Viappiani}, \bibinfo{person}{Boi
  Faltings}, {and} \bibinfo{person}{Pearl Pu}.}
  \bibinfo{year}{2006}\natexlab{}.
\newblock \showarticletitle{Preference-based search using example-critiquing
  with suggestions}.
\newblock \bibinfo{journal}{\emph{Journal of artificial intelligence Research}}
   \bibinfo{volume}{27} (\bibinfo{year}{2006}), \bibinfo{pages}{465--503}.
\newblock


\bibitem[\protect\citeauthoryear{Vig, Sen, and Riedl}{Vig
  et~al\mbox{.}}{2011}]%
        {vig2011navigating}
\bibfield{author}{\bibinfo{person}{Jesse Vig}, \bibinfo{person}{Shilad Sen},
  {and} \bibinfo{person}{John Riedl}.} \bibinfo{year}{2011}\natexlab{}.
\newblock \showarticletitle{Navigating the tag genome}. In
  \bibinfo{booktitle}{\emph{Proceedings of the 16th international conference on
  Intelligent user interfaces}}. \bibinfo{pages}{93--102}.
\newblock


\bibitem[\protect\citeauthoryear{Wu, Luo, Sanner, and Soh}{Wu
  et~al\mbox{.}}{2019}]%
        {wu2019deep}
\bibfield{author}{\bibinfo{person}{Ga Wu}, \bibinfo{person}{Kai Luo},
  \bibinfo{person}{Scott Sanner}, {and} \bibinfo{person}{Harold Soh}.}
  \bibinfo{year}{2019}\natexlab{}.
\newblock \showarticletitle{Deep language-based critiquing for recommender
  systems}. In \bibinfo{booktitle}{\emph{Proceedings of the 13th ACM Conference
  on Recommender Systems}}. \bibinfo{pages}{137--145}.
\newblock


\end{thebibliography}

\end{document}